\documentclass[published]{JHEP3} 
\usepackage{epsfig,multicol,psfrag}
\usepackage{bm}
\usepackage{amsfonts}
\usepackage{epic}
\usepackage{eepic}
\def\bea{\begin{eqnarray}}
\def\eea{\end{eqnarray}}
\def\nn{\nonumber}
\def\beq{\begin{equation}}
\def\eeq{\end{equation}}
\def\ba{\beq\new\begin{array}{c}}
\def\ea{\end{array}\eeq}

\newcommand{\A}{{\cal A}}
\newcommand{\cF}{{\cal F}}

\newcommand{\de}{ \Delta}
\newcommand{\s}{{\sigma}}

\newcommand{\lb}{\label}

\def\lb{\label}

\newcommand{\e}{{\rm e}}

\newif\ifold \oldtrue \def\new{\oldfalse}

\JHEP{00(2005)000}

\title{
   \begin{flushright} \begin{small}
     DTP-MSU/05-09, ITEP-TH-48/05, LAPTH-1105/05 \\
  \end{small} \end{flushright}
\vspace{.5cm} {\Large\bf More on general $p$-brane solutions  } }

\author
{D. Gal'tsov$^{a,d}$, S. Klevtsov$^{a,b,c}$, D. Orlov$^{a}$ and
G. Cl\'ement$^{d}$
\\ \\
{$^{a}$Department of Theoretical Physics, Moscow State University,
119899, Moscow, Russia}
\\
{$^{b}$ ITEP, 25 B. Cheremushkinskaya, Moscow 117259,
Russia}\\
{$^{c}$ Department of Physics and Astronomy,
Rutgers University, Piscataway, NJ 08854-8019, USA}\\
{$^{d}$Laboratoire de Physique Th\'eorique LAPTH (CNRS), B.P.110,
F-74941 Annecy-le-Vieux cedex, France}

E-mail: \email{galtsov@mail.phys.msu.ru},
\email{klevtsov@itep.ru}, \email{orlov\_d@mail.ru},
\email{gclement@lapp.in2p3.fr}}

\received{June xx, 2005}        
\revised{}
\accepted{}        

\abstract{Recently it was found that the complete integration of
the Einstein-dilaton-antisymmetric form equations depending on one
variable and describing static singly charged $p$-branes leads to
two and only two classes of solutions: the standard asymptotically
flat black $p$-brane and the asymptotically non-flat $p$-brane
approaching the linear dilaton background at spatial infinity.
Here we analyze this issue in more details and generalize the
corresponding uniqueness argument to the case of partially
delocalized branes. We also consider the special case of
codimension one and find, in addition to the standard domain wall,
the black  wall solution. Explicit relations between our solutions
and some recently found $p$-brane solutions ``with extra
parameters'' are presented.}


\begin{document}

\begin{center}
\rule{5cm}{1pt}
\end{center}

\bigskip
\section{Introduction}
Classical supergravity solutions  describing $p$-branes   were
extensively studied during the past decade
\cite{HoSt91,Gu92,DuLu94,LuPoSeSt95,LuPo95,LuPoXu95,DuKhLu95,DuLuPo96,LuPoSeSt96,St98,GaRy98,IvMe01,Du99}.
It is generally believed that a singly charged static p-brane
solution  depends on two parameters, the mass and the charge
(densities) of the brane (up to the dilaton value at infinity)
\cite{HoSt91,DuLu94}, these solutions are asymptotically flat and
possess a regular event horizon. Black $p$-branes have the
$ISO(p)\times R$ symmetry of the world volume (with $R$
corresponding to the time direction), which is enhanced to the
full Poincar\'e symmetry $ISO(p,1)$ in the extremal (BPS) case.
Recently the uniqueness of these solutions was investigated in
more detail \cite{ClGa04,GaLeCl04}, and it was shown that, once
the absence of naked singularities is required, the generic
solution must be either asymptotically flat (standard) or to
approach the linear dilaton asymptotic at spatial infinity
\cite{ClGaLe02,ClLe04,Le04}. In ten dimensions the latter class
corresponds to the near-horizon limit of near-extremal p-branes
and provides the holographic counterpart to the thermal phase of
the dual conformal field theories with sixteen supercharges
\cite{ItMaSoYa98,Be99, BoSkTo99,MaSt97}.

Meanwhile some apparently different claims can be encountered in the
recent literature. The complete integration of the
Einstein-dilaton-antisymmetric form system for a single brane was
performed \cite{ZhZh99} and a family of $ISO(p)\times R$ solutions
was presented, containing four free parameters. An interpretation of
one of the extra parameters was attempted in \cite{BrMaOz01} (see
also \cite{OhYok02,LuRoyDel04,KoAsMa04}): the $ISO(p,1)$ subfamily of
the solutions of \cite{ZhZh99} was treated as describing the
brane-antibrane system in the sense of Sen \cite{Se99}, the
corresponding extra degree of freedom  being associated with the
tachyon. The solutions of \cite{ZhZh99} were also invoked in some
recent attempts to find the supergravity description for stable
non-BPS branes of string theory \cite{BerVecFra00,Frau:2000gk,
Alberghi:2001fy,Bertolini:2002de}. Other generalizations of the
$ISO(p,1)$ solution were given in \cite{ChGaGu02}. Besides containing
additional parameters, the solutions presented in this paper also
describe a more general structure of the transverse space, namely,
$SO(k)\times R^{q},\; q=d-p-k-2$ (cylinder), $R^{q+k+1},\;$ as well
as the case of the hyperbolic geometry $SO(k-1,1)\times R^{q},\;$.
However,  the geometric structure of the $p$-brane type solutions
with extra parameters, in particular the nature of singularities, was
not sufficiently investigated so far.

The purpose of the present paper is to generalize the analysis of
\cite{ClGa04,GaLeCl04} to the case of a more general structure of
the transverse space, extend the uniqueness proof for partially
localized branes, consider the particular case of  domain walls,
and to clarify the relationship with other ``general'' $p$-brane
solutions. We confirm our previous uniqueness argument saying that
there are two and only two classes of solutions without naked
singularities: the standard asymptotically flat branes (black or
BPS), and black branes on the linear dilaton background.

The paper is organized as follows. In the next two section we
describe the system and its complete integration via reduction to the
separate Liouville equations. In Sec. 4 the special points of the
generic solutions are analyzed  by computing the Ricci and
Kretschmann scalars and studying the radial null geodesics. We
identify the horizons, set up the horizon regularity conditions, and
list all possibilities when singularities are hidden inside the
horizon. We find three different classes of solution one of which is
compact singular, and two other correspond to either asymptotically
flat case, or asymptotically linear dilaton backgroungd case. The
general regular asymptotically flat black solution is constructed in
Sec. 5, it contains three free parameters (mass, charge and the
asymptotic value of the dilaton) and no any 'extra parameters'. The
following Sec. 6 is devoted to the construction of the black branes
on the linear dilaton background including their thermodynamical
properties. The Sec. 7 contains an analysis of the case of the
codimension one. Here in addition to the standard supergravity domain
wall we obtain the black solution which can be interpreted also as
the black wall on the linear dilaton background. In Sec. 8 we give
explicit relations between our generic solution and some other
previously found general solutions and review them in the spirit of
our reasoning.

\section{Setup}

Consider the  standard action  containing the metric, the $q$-form
field strength, F$_{[q]}$, and the dilaton, $\phi$, coupled to the
form field with the coupling constant $a$
\bea\label{action}
S =
\int d^d x \sqrt{-g} \left( R - \frac12
\partial_\mu \phi
\partial^\mu \phi - \frac1{2\, q!} \, {\rm e}^{a\phi} \, F_{[q]}^2
\right).
\eea

The corresponding equations of motion
\bea
R_{\mu\nu} - \frac12 \partial_\mu \phi \partial_\nu \phi - \frac{{\rm e}^{a\phi}}{2(q-1)!}
\left[ F_{\mu\nu_2\cdots\nu_q}F_\nu{}^{\nu_2\cdots\nu_q}- \frac{q-1}{q(d-2)} F_{[q]}^2 \,g_{\mu\nu} \right]&=&0,\label{Ein} \\
\partial_\mu \left( \sqrt{-g} \, {\rm e}^{a\phi} \,F^{\mu\nu_2\cdots\nu_q} \right)&=&0, \label{form} \\
\frac1{\sqrt{-g}}\, \partial_\mu \left( \sqrt{-g} \partial^\mu \phi \right) - \frac{a}{2\, q!} {\rm e}^{a\phi} F_{[q]}^2&=&0,\label{dil}
\eea
  are invariant under the  discrete S-duality:
\bea
g_{\mu\nu} \to g_{\mu\nu}, \qquad  F \to \e^{-a\phi} \ast F, \qquad  \phi \to -\phi, \label{duality}
\eea
where $\ast$ denotes a $d$-dimensional Hodge dual, so  we will restrict
ourselves to purely magnetic solutions.

We consider black $p$-brane solutions  with a $p+1$ dimensional
world volume and a transverse space being the $q$ dimensional
space $\Sigma_{k,\sigma}\times\mathbb{R}^{q-k}$ ($p+q=d-2$):
\bea
ds^2 =& -& {\rm e}^{2 B} dt^2 + {\rm e}^{2 D} (dx_1^2 + \ldots + dx_p^2)
 +{\rm e}^{2 A} dr^2+ {\rm e}^{2 C} \,d\Sigma_{k,\sigma}^2+\nn\\
 &+& {\rm e}^{2 E} (dy_1^2 + \ldots +dy_{q-k}^2), \label{metric}
\eea
parametrized by five functions of the radial
variable $A(r),\,B(r),\, C(r),\,D(r)$ and $E(r)$. The space
$\Sigma_{k,\sigma}$ for $\sigma=0,+1,-1$ is a $k$-dimensional
constant curvature space --- a flat space, a sphere and a
hyperbolic space respectively:
\bea
d\Sigma_{k,\sigma}^2 = \bar g_{ab} dz^a dz^b = \left\{
 \begin{array}{ll}
 d \varphi^2 + \sinh^2\varphi \, d\Omega_{(k-1)}^2, \qquad  & \sigma=-1,\\
 d \varphi^2 + \varphi^2 \, d\Omega_{(k-1)}^2, \qquad  & \sigma=0,\\
 d \varphi^2 + \sin^2\varphi \, d\Omega_{(k-1)}^2, \qquad  & \sigma=+1,
 \end{array} \right.\label{gmetric}
\eea
satisfying \bea \bar R_{ab} = \sigma (k-1)
\bar g_{ab}. \eea  In the case of codimension one (a domain wall)
$q=k=0$, that is the $\Sigma$ and $y$- parts in (\ref{metric}) are
absent and the coordinate $r$ varies on the full line
$(-\infty,\infty)$.

The magnetically charged $p$-brane is supported by the form field
\bea \label{SolF} F_{[q]} = b  \,\,
\mbox{vol}(\Sigma_{k,\sigma})\wedge dy_1\wedge...\wedge dy_{q-k},
\eea satisfying Eq. (\ref{form}), where $b$ is the field strength
parameter, with $\mbox{vol}(\Sigma_{k,\sigma})$ denoting the
volume form of the  space $\Sigma_{k,\sigma}$.

The Ricci tensor for the metric
(\ref{metric}) has the non-vanishing components
\bea
R_{tt} &=& \e^{2(B-A)}\left[B''+B'(B'-A'+(q-k)E'+kC'+pD')\right],\\
R_{\alpha\beta} &=& -\e^{2(D-A)}\left[ D'' + D'(B'-A'+(q-k)E'+kC'+pD')\right]\delta_{\alpha\beta},\\
R_{rr} &=& -B''-B'(B'-A')-k(C''+C'^2-A'C')-\nn\\
&&-(q-k)(E''+E'^2-A'E')-p(D''+D'^2-A'D'), \label{Rrr}\\
R_{ab} &=& -\left\{\e^{2(C-A)}\left[C''+C'(B'-A'+(q-k)E'+kC'+pD')\right]\right.\nn\\
&&\left.\phantom{\e^{(A)}}-\sigma(k-1)\right\}\,\bar g_{ab}, \label{Rab}\\
R_{ij} &=& -\e^{2(E-A)}\left[E''+E'(B'-A'+(q-k)E'+kC'+pD')\right]\delta_{ij},
\eea
where  primes denote derivatives with respect to $r$.

By transforming the radial coordinate, one can fix one
of the metric functions in (\ref{metric}). This freedom can be
encoded in the following gauge function $\cF$ \bea \lb{gauge0} \ln
\cF=-A+B+kC+pD+(q-k)E. \eea Fixing  $\cF$ we thereby choose some
gauge. First we will find the general solutions without imposing
any gauge, i.e. with arbitrary $\cF$. From the Einstein equations
 we  find four equations for $B,\,C,\,D$ and $E$ with similar
differential operators
\bea
&&B''+B'\frac{\cF'}{\cF}=\frac{(q-1)b^2\e^{G}}{2(d-2)\cF^2},
\label{EqB} \\ &&C''+C'\frac{\cF'}{\cF}=
-\frac{(p+1)b^2\e^{G}}{2(d-2)\cF^2}+\sigma(k-1)\e^{2(A-C)},\label{EqC}
\\
&&D''+D'\frac{\cF'}{\cF}=\frac{(q-1)b^2\e^{G}}{2(d-2)\cF^2},\label{EqD}
\\ &&E''+E'\frac{\cF'}{\cF}=-\frac{ (p+1)b^2\e^{G}}{2(d-2)\cF^2},
\label{EqE} \eea where \bea \lb{G} G=a\phi+2B+2pD, \eea
 and the following equation
\bea
(A+\ln\cF)''&-&A'(A+\ln\cF)'+B'^2+\nn\\
&+&kC'^2+pD'^2+(q-k)E'^2+ \frac12\phi'^2 =\frac{(q-1) b^2{\rm e}^{G }}{2(d-2)\cF^2}.\eea
The dilaton equation Eq.(\ref{dil}) takes a similar form
\bea
\label{EqPhi}
\phi'' +\phi'\frac{\cF'}{\cF} = \frac{a b^2{\rm e}^{G}}{2\cF^2}.
\eea

To simplify the system we introduce a new function instead of $A$:
\bea
\A=A+\ln\cF = B + kC + pD + (q-k)E,
\eea
and a new variable $\tau$, satisfying
\bea
\label{tau} \frac{d\tau}{dr}=\frac{(k-1)}{\cF}.
\eea
Then, denoting the derivatives with
respect to $\tau$ by a dot, we obtain the following system:
\bea
\ddot B &=&\frac{(q-1)b^2}{2(k-1)^2(d-2)}\;\e^{G}, \label{EqBt} \\
\ddot C&=&-\frac{(p+1)b^2 }{2(k-1)^2 (d-2)}\;\e^{G}+\frac{\sigma}{(k-1)}\;\e^{2(\A-C)}, \label{EqCt} \\
\ddot D&=&\frac{(q-1)b^2}{2(k-1)^2 (d-2)}\;\e^{G}, \label{EqDt} \\
\ddot E&=&-\frac{(p+1)b^2}{2(k-1)^2 (d-2)}\;\e^{G}, \label{EqEt} \\
\ddot \phi&=&\frac{ab^2}{2(k-1)^2}\;\e^{G}, \label{Eqfit} \eea
and
\bea
\label{EqAt} \ddot \A -{\dot \A}^2+{\dot B}^2+k{\dot
C}^2+p{\dot D}^2+ (q-k){\dot E}^2+\frac12{\dot
\phi}^2=\frac{(q-1)b^2}{2(k-1)^2 (d-2)}\;\e^{G}.
\eea
After solving the system with respect to ${\cal A}$, this last equation becomes a
constraint equations for the other functions involved.

\section{General solution}
The integration procedure is similar to that used in
\cite{GaLeCl04}. First we observe that the functions $D,\, B$, and
the quantities $-(q-1)E/(p+1)$ and $(q-1)\phi/a(d-2)$ may differ
only by a solution of the homogeneous equation, which is a linear
function of $\tau$. Therefore
\bea
D&=&B+d_1\tau+d_0,\lb{DB}\\
E&=&-\frac{p+1}{q-1}B+e_1\tau+e_0,\lb{EB}\\
\phi&=&\frac{a(d-2)}{q-1}B+ \phi_1\tau+\phi_0,\lb{FB}
\eea
where
$d_0,\,d_1,\,e_0,\,e_1,\,\phi_0,\,\phi_1$ are free constant
parameters. Substituting this into (\ref{G}) one finds :
\bea
\lb{GB} G=\frac{\de (d-2)}{(q-1)}B+g_1\tau+g_0,
\eea
where
\bea
g_{0,1}&=&a\phi_{0,1}+2pd_{0,1},\\
\Delta&=&a^2+\frac{2(p+1)(q-1)}{d-2},
\eea
so the Eq. (\ref{EqBt})
becomes a decoupled equation for $G$
\bea
\lb{Geq}\ddot G=\frac{b^2\de }{2(k-1)^2}\;\e^{G}.
\eea
Its general solution,
depending on two integration constants $\alpha$ and $\tau_0$,
reads
\bea
\lb{SolG} G=\ln\left(\frac{\alpha^2(k-1)^2}{\de
b^2}\right)-\ln\left[\sinh^2\left(\frac{\alpha}{2}
(\tau-\tau_0)\right)\right],
\eea
with $\alpha^2 > 0$ being equal to
the first integral of Eq. (\ref{Geq})
\bea
\label{al} {\dot G}^2-\frac{b^2\de}{(k-1)^2}\;{\rm e}^{G}=\alpha^2.
\eea
Using this
we integrate the remaining equations (\ref{EqCt}, \ref{EqAt}) with
account for the definition of the gauge function (\ref{gauge0}), which
with the substitution (\ref{DB},\ref{EB}) reads \bea \lb{gage}
\A-kC=\frac{(p+1)(k-1)}{q-1}\,B+p(d_1\tau+d_0)+(q-k)(e_1\tau+e_0).
\eea  Then the linear combination
\bea
\lb{H} H=2(\A-C)
\eea
will satisfy the second decoupled Liouville equation
\bea
\lb{EqH} \ddot H=2\s \e^{H},
\eea
whose first  integral is
\bea
\label{Hint} {\dot H}^2 -4\s {\rm e}^{H}=\beta^2.
\eea
Its general solution depending on two parameters $\beta,\,\tau_1$ reads, for $\beta^2 > 0$,
\bea
\lb{SolH}
H = \left\{
\begin{array}{ll} 2\ln\frac{\beta }{2}-\ln\left( \sinh^2[\beta
(\tau-\tau_1)/2]
\right), \qquad  & \sigma=1, \\
\pm \beta (\tau-\tau_1), & \sigma=0, \\
2\ln\frac{\beta}{2}-\ln\left( \cosh^2[\beta (\tau-\tau_1)/2
] \right), & \sigma=-1, \end{array}\right.
\eea

Finally, expressing the metric functions $A,\,C$ from
(\ref{gage},\ref{H}), one can write the entire solution in terms
of $G,\,H$ as follows:
\bea
\label{solution}
A&=& \frac{k}{2(k-1)}\,H-\frac{p+1}{\de(d-2)}\,G-\ln\cF+c_1\tau+c_0,\lb{Sol1}\\
B&=&\frac{(q-1)}{\de(d-2)}\left(G-g_1\tau-g_0\right),\lb{Sol5}\\
C&=&\frac{1}{2(k-1)}\;H-\frac{p+1}{\de(d-2)}\; G+c_1\tau+c_0,\lb{Sol3}\\
D&=&\frac{(q-1)}{\de(d-2)}\left(G-g_1\tau-g_0\right)+d_1\tau+d_0,\lb{Sol2}\\
E&=&-\frac{p+1}{\de(d-2)}\left(G-g_1\tau-g_0\right)+e_1\tau+e_0,\lb{Sol4}\\
\phi&=&\frac{a}{\de}\;G+f_1\tau+f_0,\lb{Sol6} \eea
where the following linear combinations of the previously introduced
parameters are used
\bea
&&g_{0,1}=a\phi_{0,1}+2pd_{0,1},\lb{phi} \\
&&c_{0,1}=\frac{(p+1)a\phi_{0,1}}{\de(d-2)}-
\frac{p(a^2(d-2)+2(p+1)(q-k))}{\Delta(d-2)(k-1)}d_{0,1}
-\frac{(q-k)}{k-1}e_{0,1},\\
&&f_{0,1}=(1-\frac{a^2}{\Delta})\phi_{0,1}-\frac{2pa}{\de}d_{0,1}.\lb{Solc}
\eea

Our solution depends on eleven parameters:
$b,\,d_0,\,d_1,\,e_0,\,e_1,\,\phi_0,\,\phi_1,\,\tau_0,\,\tau_1,\,\alpha,\,\beta$,
which satisfy a constraint following from the Eq. (\ref{EqAt})
\bea
\label{constraint}
&&-\frac{k\beta^2}{4(k-1)}+\frac{\alpha^2}{2\Delta}+
 (k-1)c_1^2  +\left(\frac{(q-1)g_1}{\de(d-2)}\right)^2+
\nn\\
&&+p\left(d_1-\frac{g_1(q-1)}{\de(d-2)}\right)^2+
(q-k)\left(e_1+\frac{(p+1)g_1}{\Delta (d-2)}\right)^2+
\frac{1}{2}f_1^2=0.
\eea
The coordinate $\tau$ is defined by (\ref{tau}) only up to a
translation, so without loss of generality we can set
\bea \tau_1=0. \eea
Also, we can always rescale the coordinates $x_{\alpha}$ and $y_{i}$
of the flat spaces $R^p$ and $R^{q-k}$ so that
\bea d_0 = e_0 = 0. \eea
We then remain with seven independent parameters. Note that the
constraint involves only five of
these. Considering the signs of the different terms in Eq. (\ref{constraint})
we find three nontrivial cases:
  $\alpha$ and $\beta$ are both real,
  $\alpha$ is pure imaginary and $\beta$ real, and
  $\alpha$ and $\beta$ are both pure imaginary.
For imaginary values of parameters $\alpha$ and $\beta$ one should
replace $\sinh,\, \cosh$ $\rightarrow$ $\sin,\,\cos$. In what
follows we will deal mostly with the  case of real $\alpha$ and
$\beta$.

We have obtained the general solution for the three cases
$\sigma=1,0$ and $-1$ (spherical, toroidal or hyperbolic sector
$\Sigma$). One can show that there are no asymptotically flat
stationary solutions for $\sigma=0$ and $-1$.
In the remainder of the paper we restrict our
analysis to the case of $SO(k)$ symmetry, i.e. $\sigma=1$.

\section{Special points}

The solution constructed above have several special points: $\tau
=0, \tau=\tau_0, \tau=\pm\infty$. To reveal their physical meaning
consider the Ricci scalar which can be expressed using the field
equations as follows
\bea
\label{Req}
R=\frac{(k-1)^2}{2}\e^{-2{\cal A}}\left(\dot{\phi}^2+\frac{b^2(d-2q)}{(d-2)(k-1)^2}\e^G\right).
\eea
Note that for $d=2q$ and a constant dilaton (which is
consistent with the field equations if $a=0$), the scalar
curvature is identically zero, as a consequence of the conformal
invariance of the corresponding form field.

Substituting here the solution (\ref{Sol1}-\ref{Sol6}) one obtains
\bea
\label{Req1}
R &=& \frac{(k-1)^2}2 \e^{-2c_0-2c_1\tau}
\bigg[\frac{\alpha(k-1)}{\sqrt{\Delta}b}\bigg]^{4\frac{p+1}{\Delta(d-2)}}
\bigg[\frac{2\sinh(\frac{\beta}2\tau)}{\beta}\bigg]^{\frac{2k}{k-1}}
\bigg[\sinh\bigg(\frac{\alpha}
2(\tau-\tau_0)\bigg)\bigg]^{-4\frac{p+1}{\Delta(d-2)}-2} \times \nn \\
&&\times\bigg\{\bigg(f_1\sinh\bigg(\frac\alpha
2(\tau-\tau_0)\bigg)-\alpha\frac{a}{\Delta}\cosh\bigg(\frac{\alpha}2
(\tau-\tau_0)\bigg)\bigg)^2
+\alpha^2\frac{d-2q}{\Delta(d-2)}\bigg\}. \eea

An additional information about singularities can be found from
the square of the Riemann tensor (the Kretschmann scalar)
$K=R_{\alpha\beta\gamma\delta}R^{\alpha\beta\gamma\delta}$. To
present it in a compact notation, we introduce the four-component
quantities $Y_i=\{B,D,C,E\},l_i=\{1,p,k,q-k\}, i=1, 2, 3, 4$. Then
the Kretschmann scalar takes the following form:
\bea
\label{Keq}
K&=& 4(k-1)^4\e^{-4{\cal
A}}\biggl(\sum_{i}l_i(-\ddot{Y}_i-\dot{Y_i}^2+ \dot{\cal
A}\dot{Y_i})^2+\frac12\sum_{i\neq j}l_il_j\dot{Y}_i^2\dot{Y}_j^2+ \nn\\
&&+\frac12\sum_{i}l_i(l_i-1)\dot{Y}_i^4-
\sigma\frac{k}{k-1}\dot{C}^2\e^{2({\cal A}-C)}+
\frac12\frac{k}{(k-1)^3} \e^{4({\cal A}-C)}\biggr).
\eea

From these formulas one can see that $\tau=\tau_0$ generically
is the singular point of the geometry unless $\tau_0=0$, in which
case the singularity can be avoided for some choice of parameters. One
of the points $\tau=\pm \infty$ generically is also singular.
Namely, if the parameters in the exponential are chosen so that
one of them is regular, another will be singular and vice versa.
Another special feature associated with these points is that the
the metric coefficient
\bea
g_{tt}=\e^{2B}=
\bigg[{\displaystyle \frac{\alpha^2(k-1)^2\e^{-(g_0+g_1\tau)}}
{\Delta b^2\sinh^2(\alpha(\tau-\tau_0)/2)}}\bigg]^{\frac{2(q-1)}
{\Delta(d-2)}}
\eea
may vanish there, corresponding to horizons of the geometry.
This may happen  for \bea
\tau=+\infty,\,\,\,\mbox{if}\,\,\,|\alpha|>-g_1
\eea
and
\bea
\tau=-\infty,\,\,\,\mbox{if}\,\,\,|\alpha|>g_1.
\eea
In what follows we will choose the second option $\tau=-\infty$ for the
location of the regular event horizon, in which case the point
$\tau=+\infty$  generically will be singular (unless some special
values of the parameters are chosen).

Further information about the geometrical structure of the solution
can be extracted from the analysis of radial geodesics $x^\mu
(\lambda)=(t(\lambda), r(\lambda))$
\bea
t''+2\frac{dB}{dr}r't'=0,\nn\\
r''+\frac{dA}{dr}r'^2 + \e^{2(B-A)}\frac{dB}{dr}t'^2=0,
\eea
where primes denote the derivatives with respect to the affine parameter
$\lambda$. The first integral of this
system reads
\bea
t'=\e^{-2B+C_1},r'=\pm \e^{C_1}\sqrt{\e^{-2(A+B)} + C_2\e^{-2(A+C_1)}},\label{fd}
\eea
where $C_{1,2}$ are integration constants. Null radial geodesics correspond
to $C_2 = 0$, in which case the second equation reduces (after a
rescaling of the affine parameter) to
\bea r'=\frac1{k-1}\e^{-(A+B)}. \eea
 Passing to our new variable $\tau$, related to $r$ via $d\tau/dr=(k-1)/\cal F$,
we can rewrite the null radial geodesic equation as
\bea \tau'= \e^{-({\cal A}+B)},\eea
so that
\bea\lb{afftau} \lambda\sim\int \e^{{\cal A}+B}d\tau. \eea

We wish to identify the values $\tau=\tau_\infty$ which may
correspond to points at an infinite distance from the horizon. For
this we evaluate the integrals in the vicinity of $\tau_\infty$
obtaining the leading term
\bea
\lambda\sim \int (\tau-\tau_\infty)^{\gamma-1}d\tau \sim
(\tau-\tau_\infty)^{\gamma}.
\eea
If $\gamma<0$, the affine distance to this point will be infinite. We have three particular
cases:
\bea
&1)&\tau_0\neq0,\qquad\tau_\infty=0,\qquad\gamma=-\frac1{k-1}<0 \lb{case1},\\
&2)&\tau_0\neq0,\qquad\tau_\infty=\tau_0,\qquad\gamma=\frac{a^2(d-2)+2(pq+1)}{\Delta(d-2)} >0,\lb{case2}\\
&3)&\tau_0=0,\qquad\tau_\infty=0,\nn\\
&&\gamma=-\frac{a^2(d-2)+2(q-1)(k-1)+2(p+1)(q-k)}{\Delta(d-2)(k-1)}<0.\lb{case3} \eea
Thus, $\tau=0$ is an infinitely distant point in
both cases $\tau_0=0$ and $\tau_0\neq 0$. One can further show
that in the case $\tau_0\neq 0$ both the Ricci and Kretschmann
scalars vanish at $\tau=0$; we will see that in this case
the metric is asymptotically flat. In the second case
(\ref{case2}), the non-zero $\tau_0$, which is a
singular point of the solution, is located at a finite
affine distance, thus representing a true singularity.

Now let us investigate the regularity of the event
horizon $\tau=-\infty$. Near a regular horizon  the metric
coefficient $\e^{2B}$  must behave \cite{GaLeCl04} as \bea
\label{hor2} \e^{2B}\sim\lambda^n, \eea where $\lambda=0$
corresponds to the position of the horizon, and the integer $n$ is
equal to unity for a  non-degenerate horizon, and $n\geq 2$ for a
degenerate horizon. Differentiating (\ref{hor2}) we obtain the
following condition \bea \e^{-({\cal
A}+\frac{3n-2}{n}B)}\frac{d}{d\tau}\e^{2B}\sim \mbox{const} \eea at
the horizon. For the solution (\ref{Sol1}-\ref{Sol6}) we get
\bea \lb{horAB} \e^\A\sim \e^{\frac{2-n}{n}B}.\eea
Taking into account this equation we can rewrite the equation
(\ref{EqAt}) near the horizon as
\bea
\label{ceq}
\frac{4(n-1)}{n^2}\dot{B}^2+k\dot{C}^2+p\dot{D}^2+(q-k)\dot{E}^2+\frac12\dot{\phi}^2 \sim 0.
\eea
The left-hand side being a sum of squares, all of them must vanish, so
that we obtain the
following conditions on the behavior of the metric exponents at
the horizon for $n=1$:
\bea \dot C=\dot D=\dot E=\dot\phi=0,\quad \mbox{at} \quad \tau=-\infty, \eea
which are equivalent for the solution (\ref{Sol1}-\ref{Sol6}) to the following conditions on
the free parameters:
\bea \label{parhor} |\alpha|=|\beta|=-2d_1=\frac{2(q-1)}{p+1}e_1=-\frac{2(q-1)}{a(d-2)}\phi_1. \eea
 In what follows without loss of generality we will choose $\alpha$ to be non-negative. The
calculation shows that under these conditions the constraint (\ref{constraint}) is satisfied automatically.

In the degenerate case $n\geq 2$, we obtain from (\ref{ceq})
the additional condition
\bea
\dot{B}=0,\quad \mbox{at}\quad  \tau=-\infty.
\eea
Combined with the conditions
(\ref{parhor}), this leads to
\bea \alpha=\beta=d_1=\phi_1=e_1=0. \eea
Note that in the case $\alpha = \beta = 0$, (\ref{SolG}) and
(\ref{SolH}) should be replaced by
\bea
G= -2\ln|\tau-\tau_0| + \ln(4(k-1)^2/b^2\Delta), \qquad H = -2\ln|\tau|,
\eea
leading to the behavior near the horizon $\tau \to -\infty$
\bea
\A \sim \mu\ln|\tau|, \qquad B \sim \nu\ln|\tau|,
\eea
with
\bea
\mu = \frac{2(p+1)}{\Delta(d-2)}- \frac{k}{k-1}, \qquad \nu = -\frac{2(q-1)}{\Delta(d-2)}.
\eea
From (\ref{afftau}), $d\lambda \sim \tau^{\mu+\nu}d\tau$,
leading to
\bea
\e^{2B} \sim \tau^{2\nu} \sim \lambda^{2\nu/(\mu+\nu+1)}.
\eea
This is consistent with $\e^{2B} \sim \lambda^2$ only if $\mu + 1 = 0$,
leading to the conditions for the existence of regular degenerate horizons
\bea
 a^2(d-2) + 2(p+1)(q-k) = 0 \quad\Leftrightarrow\quad (a^2 = 0\,,\;\;q = k).
\eea

Coming back to the non-degenerate case, we note that once the
position of the regular event horizon is chosen as $\tau=-\infty$,
the Ricci scalar will be finite there, while at $\tau=+\infty$ (an
inner horizon) it will diverge as \bea R\sim \left(a\e^{\alpha\tau}
+{\rm const}\right)
\e^{\left[\frac{2(a^2(p+q)+2(p+1)(q-k))}{(q-1)\Delta(p+q)}\right]
\alpha\tau}, \eea since the coefficient in the exponential is
non-negative (for our choice $\alpha\geq 0$). In the special case
where the coefficient vanishes, i.e. $a=0,\, q=k$, the inner horizon
is apparently regular (obviously, the degenerate horizon discussed above will
arise when this inner horizon coincides with the outer
horizon). However, a closer look reveals that the Kretschmann scalar
(\ref{Keq}), which for $a=0,\, q=k$ diverges at $\tau=\infty$ as
\bea
K \sim \alpha^4(k-1)^4\frac{p^2(p-1)}{(p+1)^3}\,\e^{2\alpha\tau}
\eea
can be finite only for $p=0$ or $p=1$ (see next section).

Let us finally comment on the possibility of imaginary values of the
parameters $\alpha$ and $\beta$. As previously mentioned, for
imaginary values of  $\alpha$  one should
replace in the solution the hyperbolic functions $\sinh,\, \cosh$ by
the trigonometric functions  $\sin,\,\cos$. In this case we will have
an infinite sequence of singularities, located at the points
$\tau=\tau_0+2\pi n/|\alpha|$ ($n$ integer), together with an
infinite sequence of points at infinity $\tau=2\pi m/|\alpha|$
($m$ integer) if $\beta$ is also imaginary. In any sector
between two consecutive singularities (or between a singularity and
the next point at infinity), the metric function is bounded,
so that no horizon can occur. In what follows we restrict ourselves to
real values of these parameters.

\section{The black solutions in Schwarzschild-like coordinates}

As discussed in the previous section, the nondegenerate black solution
(\ref{Sol1}-\ref{Sol6}) with $\tau_1 = 0$ and conditions (\ref{parhor})
has a horizon at $\rho\to -\infty$, a point at infinity at $\tau = 0$,
and a singularity at $\tau = \tau_0$ if $\tau_0 \neq 0$. According to
the sign of the integration constant $\tau_0$, there are two black solution
branches:

a) \underline{$\tau_0 > 0$} At the end-point $\tau = 0$, the
solution is asymptotically flat. We will check in the following that
this can be gauge transformed to the standard black brane solution.

b) \underline{$\tau_0 < 0$} In this case the spacetime ends at the
point singularity $\tau = \tau_0$, a ``bag-of-gold''-like black
solution.

At the boundary between these two cases, \underline{$\tau_0 = 0$},
lies a critical solution which is geodesically complete outside the
horizon, but is not asymptotically flat, generalizing the black brane
solution with linear dilaton asymptotics of \cite{ClGa04}.

With a view to transform these solutions to a Schwarzschild-like
gauge, it is convenient to map the coordinate $\tau < 0$ to a new
radial coordinate $\xi$, such that the horizon $\tau \to -\infty$
maps to a finite value $\xi = \xi_+$, and the asymptotic infinity
$\tau=0$ maps to $\xi=+\infty$. This is achieved by the map
\bea
\lb{tauxi} \e^{\alpha\tau} = {\displaystyle
\frac{\xi-\xi_+}{\xi-\xi_-}},
\eea
with $\xi>\xi_+>\xi_-$. To
extend this to the full line of $\xi$ one has to consider complex
$\tau$ as shown on Fig. 1. This map defines $\xi$ (and its special
values $\xi_{\pm}$) only up to linear transformations. We first
fix the common scale of $\xi$, $\xi_+$ and $\xi_-$ by choosing
\bea
\lb{axipm} \xi_+-\xi_- = \alpha.
\eea
The image of $\tau_0$ by
the map (\ref{tauxi}) is $\xi_0$: \bea \lb{tau0}
\e^{\alpha\tau_0}={\displaystyle \frac{\xi_0-\xi_+}{\xi_0-\xi_-}}.
\eea Introducing the notation \bea \label{8}
f_\pm(\xi)=1-\frac{\xi_\pm}{\xi}, \qquad
f_0(\xi)=1-\frac{\xi_0}{\xi}, \eea we can express the functions $G$
and $H$ as \bea
H&=&\ln(\xi^2f_+f_-),\\
G&=&\ln\left(\frac{f_+f_-}{f_0^2}\right)+G_0,
\eea
with
\bea
\e^{G_0}=\frac{\alpha^2(k-1)^2}{\Delta b^2}
\sinh^{-2}\left(\frac{\alpha\tau_0}{2}\right)=
\frac{4(k-1)^2(\xi_+-\xi_0)(\xi_--\xi_0)}{\Delta b^2},
\eea
and present the generic black solution as follows
\bea
&&\e^{2\A}=|\xi|^{\frac{2k}{k-1}}f_+
f_-^{\frac{2k}{k-1}-\frac{4(p+1)}{\Delta(d-2)}}
\chi_0^{\frac{2(p+1)}{\Delta(d-2)}},\nn\\
&&\e^{2B}=f_+f_-^{-1+\frac{4(q-1)}{\Delta(d-2)}}
\chi_0^{-\frac{2(q-1)}{\Delta(d-2)}},\nn\\
&&\e^{2C}=|\xi|^{\frac2{k-1}}f_-^{\frac{2}{k-1}-\frac{4(p+1)}{\Delta(d-2)}}
\chi_0^{\frac{2(p+1)}{\Delta(d-2)}},\nn\\
&&\e^{2D}=f_-^{\frac{4(q-1)}{\Delta(d-2)}}
\chi_0^{-\frac{2(q-1)}{\Delta(d-2)}},\nn\\
&&\e^{2E}=f_-^{-\frac{4(p+1)}{\Delta(d-2)}}
\chi_0^{\frac{2(p+1)}{\Delta(d-2)}},\label{conform}\\
&&\e^{a\phi}=\e^{a\phi_0}
f_-^{\frac{2a^2}{\Delta}}\chi_0^{\frac{-a^2}{\Delta}},\\
&&F_{[q]}=2(k-1)\sqrt{\frac{(\xi_+-\xi_0)(\xi_--\xi_0)}{\Delta}}
\e^{-G_0/2}\, \mbox{vol}(\Sigma_{k,1})\wedge
dy_1\wedge...\wedge dy_{q-k},
\eea
with
\bea \lb{chi0} \chi_0(\xi) = \e^{a\phi_0-G_0}f_0^2(\xi) \eea
(recall that we have chosen $d_0= e_0=0$, so that $g_0=a\phi_0$).
The curvature scalar reads, in these coordinates,
\bea
\lb{curv}
R&=&\frac{2(k-1)^2}{\Delta}|\xi|^{-\frac{2k}{k-1}}\chi_0^{-\frac{2(p+1)}{\Delta
(d-2)}} f_-^{-\frac{2}{k-1}+\frac{4(p+1)}{\Delta (d-2)}}f_0^{-2}\times\nn\\
&&\times\left[\frac{d-2q}{d-2}(\xi_+-\xi_0)(\xi_--\xi_0)
+\frac{a^2}{\Delta}(\xi_--\xi_0)^2\frac{f_+}{f_-}
\right].
\eea

Consider first the asymptotically flat case $\tau_0 >0$. In this case,
one can choose Minkowskian  coordinates at $\tau=0$ by imposing
the conditions \bea A,\,B,\,D,\,E\rightarrow 0,\qquad  C\rightarrow
\ln r, \eea for $r\rightarrow\infty$, where $r$ is our original
radial coordinate, related to $\tau$ by (\ref{tau}). We find that the
functions $B$, $D$ and $E$ all vanish at infinity provided
\bea\lb{aphig} a\phi_0 = G_0. \eea
From the behavior
of the functions $\cal{A}$ and $C$ one can recover the asymptotic
behavior of the gauge function $\cF$ corresponding to this
solution,
\bea \label{tauinfty} \ln\cF=k\ln r,\qquad \tau=-r^{-(k-1)}. \eea
This asymptotic behavior is consistent with our coordinate
transformation (\ref{tauxi}) provided
\bea \lb{xir} \xi=r^{k-1}, \eea
whch corresponds to fixing the following gauge function
\bea\lb{gauge} \cF=r\xi f_+f_-. \eea

Now we consider the generic case $\tau_0 \neq 0$, with both signs of
$\tau_0$ allowed. We recall that the map (\ref{tauxi}) together with the
relation (\ref{axipm}) define $\xi$, $\xi_+$ and $\xi_-$ only up to a
common additive constant. Without loss of generality, we can choose
this constant so that the image of $\tau_0$ is
\bea\lb{xi00} \xi_0 = 0, \eea
leading to $f_0 = 1$. With the choice (\ref{aphig}), the generic
$\tau_0\neq0$ solution is
then obtained by setting $\chi_0 = 1$ in (\ref{conform}):
\bea
\lb{genb}
ds^2&=&-f_+f_-^{-1+\frac{4(q-1)}{\Delta(d-2)}}dt^2+
f_-^{\frac{4(q-1)}{\Delta(d-2)}}(dx_1^2+\ldots+dx_p^2)+\nn\\
&&+f_+^{-1}f_-^{-1+\frac{2}{k-1}-\frac{4(p+1)}{\Delta(d-2)}}dr^2
+r^2f_-^{\frac{2}{k-1}-\frac{4(p+1)}{\Delta(d-2)}}\,d\Omega_{k}^2+\nn\\
&&+f_-^{-\frac{4(p+1)}{\Delta(d-2)}}(dy_1^2 +\ldots+dy_{q-k}^2),\\
\e^{a\phi}&=&\e^{a\phi_0}f_-^{\frac{2a^2}{\Delta}}, \\
F_{[q]}&=&2(k-1)\frac{(r_+r_-)^{\frac{(k-1)}{2}}}{\sqrt{\Delta}}
\e^{-\frac{a}{2}\phi_0}\,\mbox{vol}(\Sigma_{k,1})\wedge
dy_1\wedge...\wedge dy_{q-k}.
\eea

The coordinate $r$ in (\ref{conform}) and (\ref{genb}) is related to
$\xi$ by
\bea \lb{rxi} r = |\xi|^{1/(k-1)}, \eea
so as to accommodate
both signs of $\xi$. For $\tau_0 > 0$,
\bea 0 < \xi_- < \xi_+ \eea
($\xi > 0$), this is the standard black brane solution \cite{DuLu94},
with a spacelike central singularity at $\xi = \xi_-$, except in
the cases ($a=0$, $q=k$, $p=0$ or 1) where the central singularity
is located at $\xi=0$ and is timelike. In both cases the
singularities are hidden behind the event horizon $\xi=\xi_+$. For
$\tau_0 < 0$, \bea \xi_- < \xi_+ < 0 \eea ($\xi < 0$), there are
generically two central singularities, a timelike singularity at
$\xi = 0$ separated from the (generic) spacelike singularity at
$\xi = \xi_-$ by the horizon at $\xi = \xi_+$.

We discuss briefly the cases $a=0$, $q = k$, and $p = 0$ or 1 where
the inner horizon $\xi = \xi_-$ becomes regular. For $p = 0$, the solution
(\ref{genb}) reduces to the Reissner-Nordstr\"om-like form
\bea
ds^2 = -f_+f_-\,dt^2 + f_+^{-1}f_-^{-1}\,dr^2 + r^2\,d\Omega_k^2.
\eea
Clearly, this is invariant under the involution (preserving the order
$\xi_- < \xi_+$)
\bea \lb{invo} \xi \to -\xi,\qquad \xi_{\pm} \to -\xi_{\mp},\eea
which transforms the $\tau_0 < 0$ solution into the asymptotically
flat $\tau > 0$ solution. For $p = 1$, the solution (\ref{genb})
reduces to
\bea
\lb{regp1}
ds^2 = -f_+\,dt^2 + f_-\,dx^2 + f_+^{-1}f_-^{-1}\,dr^2 + r^2\,d\Omega_k^2.
\eea
For $\tau_0 > 0$ ($\xi_+ > \xi_- > 0$), this is stationary in both
sectors $\xi>\xi_+$ and $0<\xi,\xi_-$, however in the inner stationary sector
the role of time is taken up by the coordinate $x$. The solution
for $\tau_0 < 0$ ($\xi_+ > \xi_- > 0$) is transformed by the involution
(\ref{invo}) into
\bea
ds^2 = -f_-\,dt^2 + f_+\,dx^2 + f_+^{-1}f_-^{-1}\,dr^2 + r^2\,d\Omega_k^2,
\eea
which is just the asymptotically flat solution (\ref{regp1}) with
$\xi_+$ and $\xi_-$ exchanged. When $\xi$ decreases from $+\infty$, first
the outer horizon $\xi=\xi_+$ is crossed, with $g_{tt}$ keeping its sign, but
$g_{xx}$ and $g_{rr}$ both flipping signs, so that in the intermediate sector
between the two horizons there are three time coordinates\footnote{In
the event that the range of the coordinate $x$ is compact, the region
$\xi<\xi_+$ will contain closed timelike curves.}. Then the
inner horizon $\xi=\xi_-$ is crossed, leading to the inner sector with
only one time coordinate $x$, and the central singularity $\xi = 0$.

Returning to the general case, we recall that the generic solution
(\ref{Sol1}-\ref{Sol6}) had seven independent parameters. We imposed
three relations (\ref{parhor}) from the condition of regularity of
the horizon (the fourth relation (\ref{parhor}) follows from the
constraint (\ref{constraint})), and the relations (\ref{aphig})
to ensure the asymptotic Minkowskian behavior. The coordinate
transformation (\ref{tauxi}) introduced two more parameters $\xi_\pm$,
corresponding to the location of the horizons, which were then
constrained by the relations (\ref{axipm}) and (\ref{xi00}). Therefore
the solution (\ref{genb}) depends on $7-4=3$ independent parameters:
$\xi_+,\,\xi_-$ and $\phi_0$. This last parameter corresponds to the
value of the
dilaton at infinity. The other two parameters enter the expressions
for the mass and charge of the solution. Indeed, following
\cite{Lu,Argurio} we can define the ADM mass per unit volume of
the brane by
\bea
\lb{mass}
\frac{M}{\mbox{vol(p-brane)}}&=&2\Omega_{k}L_{q-k}r^{k}\times\nn\\
&&\times\left.\left[\frac{k}{r}(\e^A-\e^{C-\ln r})-pD'-
(q-k)E'-k(C-\ln r)'\right]\right|_{r\rightarrow\infty},
\eea
where the prime denotes $r$-derivative, $\Omega_k$ is the volume of the unit
$k$-sphere, $L_{q-k}$ is the volume of the flat part of the transverse
space of dimension $q-k$ and we have set the Newton constant $G_d=1/16\pi$.

The ADM charge (normalized in a similar way) is defined up to the
normalization factor by (\ref{SolF}). For the solution
(\ref{genb}) we get
\bea
\lb{mass1}
\frac{M}{\mbox{vol(p-brane)}}&=&\Omega_kL_{q-k}\left[k(\xi_+-\xi_-) + \frac{4(k-1)}{\Delta}\xi_-\right],\\
\frac{P}{\mbox{vol(p-brane)}}&=&2(k-1)\Omega_kL_{q-k} \sqrt{\frac{\xi_+\xi_-}{\Delta}} \e^{-\frac{a}{2}\phi_0}.
\eea
The calculation of the entropy per unit volume of
the brane and temperature of the solution along
the lines described in \cite{Argurio} gives
\bea
T&=&\frac{k-1}{4\pi}|\xi_+|^{-\frac{2}{\Delta}}
(\xi_+-\xi_-)^{-\frac1{k-1}+\frac2{\Delta}},\\
\frac{S}{\mbox{vol(p-brane)}}&=&4\pi\Omega_kL_{q-k}
|\xi_+|^{\frac{2}{\Delta}}
(\xi_+-\xi_-)^{\frac{k}{k-1}-\frac2{\Delta}}.
\eea
The magnetic
potential $W$ is obtained by computing the dual electric form
\bea
\tilde{F}_{[p+2]} = \e^{a\phi}\star F_{[q]} = dW_{[p+1]},
\eea
with
\bea
W_{[p+1]}= W dt\wedge dx_1\wedge...\wedge dx_p,\qquad
W=2\frac{\left(\xi_+\xi_-\right)^{\frac12}}{\sqrt{\Delta}}|\xi|^{-1} \e^{\frac{a}{2}\phi_0}.
\eea
In accordance with the first law of
thermodynamics for black branes, the following identity holds \bea
dM=TdS+W|_{\xi=\xi_+}dP. \eea

\section{The critical solution $\tau_0=0$: black branes with linear
dilaton asymptotics}

From (\ref{tau0}), we see that the value $\tau_0=0$
corresponds to the limit $\xi_0 \to \pm\infty$ (see Fig. 2). In this limit, the
function $\chi_0$ defined by (\ref{chi0}) becomes
\bea
\chi_0(\xi) = \frac{\Delta b^2\e^{a\phi_0}}{4(k-1)^2}\xi^{-2}.
\eea
We note that this function enters the generic solution (\ref{conform})
through the product
\bea
f_-^2\chi_0^{-1} = \frac{4(k-1)^2}{\Delta b^2\e^{a\phi_0}}(\xi-\xi_-)^{2}.
\eea
Choosing now (without loss of generality)
\bea \xi_- = 0, \eea
fixing again the gauge function (\ref{gauge}), which gives $\xi = r^{k-1}$, and putting
\bea
\lb{r0c}
\frac{\Delta b^2\e^{a\phi_0}}{4(k-1)^2} = r_0^{2(k-1)}, \qquad \xi_+ = c,
\eea
we find that the solution
reduces to the black $p$-brane in the linear dilaton background
\cite{ClGa04} delocalized in part of the transverse space of
dimension $q-k$
\bea
ds^2&=&\left(\frac r{r_0}\right)^{\frac{4(q-1)(k-1)}{\Delta
(d-2)}}\left(-\left(1-\frac c{r^{k-1}}\right)dt^2+dx_1^2+\ldots+dx_p^2\right)+\nn\\
&&+\left(\frac {r_0}r\right)^{\frac{4(p+1)(k-1)}{\Delta
(d-2)}}\left(\left(1-\frac c{r^{k-1}}\right)^{-1}dr^2+r^2d\Omega_k^2 + dy_1^2+\ldots+dy_{q-k}^2\right), \\
\e^{a\phi}&=&\e^{a\phi_0}\left(\frac r{r_0}\right)^{\frac{2(k-1)a^2}{\Delta}},\\
F_{[q]}&=&\frac{2(k-1)\e^{-a\phi_0/2}r_0^{k-1}}{\sqrt\Delta} \,\,
\mbox{vol}(\Sigma_{k,1})\wedge dy_1\wedge...\wedge dy_{q-k},
\eea
where the radial coordinate $r$ is again related to $\xi$ by (\ref{xir}).

This solution is asymptotically nonsingular, as can be seen from
the following expression for the curvature scalar
\bea
R=\frac{4(p+1)(k-1)^2}{\Delta^2(d-2)r^2}\left(\frac
r{r_0}\right)^{\frac{4(p+1)(k-1)}{\Delta
(d-2)}}\left(\Delta-(q-1)-\frac{(d-2)a^2c}{2(p+1)r^{k-1}}\right),
\eea since \bea  2(p+1)(k-1) -\Delta (d-2)= -\left(a^2(d-2)+2
(p+1)(q-k) \right)<0.
\eea

Following Ref.\cite{ClGa04,BY,HaHo95,ChNe98,Bo00} one can get the
following expression for the mass per volume of the p-brane
\bea
\frac{M}{\mbox{vol(p-brane)}}=2\int_{\partial S(r=\infty)}\e^B(K-K_0)\sqrt{\sigma}d^{d-2}x,
\eea
where
$K=-\sigma^{\mu\nu}D_{\nu}n_{\mu}$ is the trace of the extrinsic
curvature of the boundary, $K_0=K|_{c=0}$ is its background value,
$D_\nu$ is the covariant derivative with respect to the metric
$h_{\mu\nu}=g_{\mu\nu}+u_{\mu}u_{\nu}$ induced on the space-like
hypersurface $S$ of dimension $d-1$, $u^{\mu}$ is the normal vector
to $S,\;\sigma_{\mu\nu}=h_{\mu\nu}-n_{\mu}n_{\nu}$ is the metric
induced on the boundary of  $S$ with the normal vector $n^{\mu}$.

The mass corresponding to the solution obtained reads:
\bea
\frac{M}{\mbox{vol(p-brane)}}=\Omega_kL_{q-k}\left( k-\frac{2(k-1)}{\Delta}\right)c\,.
\eea The entropy and the temperature of the brane can be
calculated as usual (see e.g. the Ref. \cite{Argurio}) giving
\bea
T&=&\frac{(k-1)}{4\pi}c^{\frac2{\Delta}-\frac1{k-1}}r_0^{-\frac{2(k-1)}{\Delta}},\\
\frac{S}{\mbox{vol(p-brane)}}&=&4\pi c^{-\frac2{\Delta}+\frac k{k-1}}r_0^{\frac{2(k-1)}{\Delta}}\Omega_kL_{q-k} .
\eea

According to the first law of thermodynamics for black branes the
following identity should hold \bea dM=TdS. \eea Applying this to
the $p$-brane with the linear dilaton asymptotic we get
\bea
\frac{d{M}}{\mbox{vol(p-brane)}}&=&\Omega_kL_{q-k}\left(-\frac{2(k-1)}{\Delta}+k\right)dc,\\
\frac{TdS}{\mbox{vol(p-brane)}}&=&\Omega_kL_{q-k}\left\{\left(-\frac{2(k-1)}{\Delta
}+k\right)dc+\frac{2(k-1)^2\Delta}c\frac{dr_0}{r_0}\right\}.
\eea

Therefore the first law of thermodynamics is true only if we do not
vary the parameter $r_0$, proportional from (\ref{r0c}) to the charge
$b$. This is due to the fact that the charge is not associated with
the black brane, but rather with the linear dilaton background
\cite{ClGa04}.

In the case of zero dilaton coupling  $a=0$ the metric takes the
form
\bea
ds^2&=&\left(\frac r{r_0}\right)^{\frac{2(k-1)}{(p+1)}}
\left(-\left(1-\frac c{r^{k-1}}\right)dt^2+dx_1^2+\ldots+dx_p^2\right)+\nn\\
&&+\left(\frac {r_0}r\right)^{\frac{2(k-1)}{(q-1)}}
\left(\left(1-\frac c{r^{k-1}}\right)^{-1}dr^2+r^2d\Omega_k^2
+dy_1^2+\ldots+dy_{q-k}^2\right),
\eea
while the curvature scalar is
\bea
R=\frac{(k-1)^2(p-q+2)}{(q-1)(p+1)r_0^{\frac{2(k-1)}{q-1}}}r^{-\frac{2(q-k)}{q-1}}.
\eea
For $q\neq k$ we have a singularity at $r=0$ and vanishing of
$R$ at infinity. Asymptotically, as $r\rightarrow\infty$, the
metric becomes:
\bea
ds^2&=&\left(\frac {\tilde{r}}{r_0}\right)^{\frac{2(k-1)(q-1)}{(p+1)(q-k)}}
\left(-dt^2+dx_1^2+\ldots+dx_p^2\right)+\left(d\tilde{r}^2+\tilde{r}^2d\Sigma_k^2\right)+\nn\\
&&+\left(\frac{r_0}{\tilde{r}}\right)^{\frac{2(k-1)}{(q-k)}}\left(dy_1^2+\ldots+dy_{q-k}^2\right),
\eea
in terms of the new radial coordinate
\bea \tilde{r}=\left(\frac r{r_0}\right)^{\frac{q-k}{q-1}}r_0. \eea

\section{Domain Walls}

Our general analysis is not valid in the special case of codimension
one, i.e., for domain walls. Consider now the domain wall for
$p=d-2,\,k=q=0$. The metric has the form \bea
ds^2=-e^{2B}dt^2+e^{2D}\left(dx_1^2+\ldots+dx_p^2\right)+e^{2A}dy^2, \eea where
the coordinate $y$ now varies on the full axis. In the magnetic
sector, we have the trivial solution for the zero-form :
\bea F_{[0]}=b={\rm const}. \eea Note that the term $F^2$ enters
the action as a cosmological constant $\Lambda=-b^2/2$, and it is
common to consider both signs for it. Accordingly, in the following we
will consider $b^2$ real. Also, the parameter \bea
\Delta=a^2-2\frac{p+1}p\eea now can be positive, negative or zero.
The equations
of motion now read
\bea
\ddot{B}&=&-\frac{b^2}{2p}e^{G},\nn\\
\ddot{D}&=&-\frac{b^2}{2p}e^{G},\nn\\
\ddot{\phi}&=&\frac{ab^2}{2}e^{G},\eea where \bea
G=2B+2pD+a\phi,\eea
and
\bea \ddot\A - \dot\A^2 + \dot{B}^2 +
p\dot{D}^2 + \frac12\dot\phi^2 = -\frac{b^2}{2p}e^{G}. \eea
Summing up  the   equations with the corresponding coefficients,
we get the following equation for $G$ \bea \label{GDW} \ddot{G}=
\frac{b^2\Delta}2e^{G}. \eea For $\Delta b^2 > 0$, this has the
solution parametrized by $\alpha \ge 0$:\bea\lb{Gwall}
G=\ln\left[\frac{\alpha^2}{\Delta b^2}\right]
-\ln\left[\sinh^2\left(\frac{\alpha}2(\tau-\tau_0)\right)\right].
\eea   In the case $\Delta=0$ ($a^2=2(p+1)/p$), the solution of
(\ref{GDW}) is a linear function : \bea G=\alpha(\tau-\tau_0).
\eea In what follows we set $\tau_0=0$ without loss of generality,
and assume $$\Delta b^2 \ge 0$$ (for $\Delta b^2 < 0$, there are
no physically interesting solutions).

In terms of  $G$  the solution for $\Delta\neq0$ reads
\bea
B&=&-\frac1{\Delta p}G + b_1\tau + b_0,\nn\\
D&=&B+d_1\tau+d_0,\nn\\
\phi&=&\frac{a}{\Delta}G+f_1\tau+f_0,\nn\\
\A&=&(p+1)B+p(d_1\tau+d_0),
\eea
where
${\cal A}=A+\ln{\cal F}$,
with the relation between the integration constants \bea
&&b_{0,1}=-\frac1{2(p+1)}[2pd_{0,1} +af_{0,1}]. \eea Half of the
constants are subject to the constraint following from the Eq.
(\ref{EqAt}): \bea\label{ceqDW}
\frac{\alpha^2}{2\Delta}+\frac{p}{p+1}(d_1^2-\Delta\frac{f_1^2}4)=0.
\eea

If $\Delta=0$, the solution is: \bea
B&=&-\frac{b^2}{2\alpha^2p}e^{\alpha\tau}+b_1\tau+b_0,\\
D&=&-\frac{b^2}{2\alpha^2p}e^{\alpha\tau}+d_1\tau+d_0,\\
\phi&=&\frac{ab^2}{2\alpha^2}e^{\alpha\tau}+f_1\tau+f_0,\\
\A&=&-\frac{b^2(p+1)}{2\alpha^2p}e^{\alpha\tau}+(b_1+pd_1)\tau+(b_0+pd_0),
\eea with the parameters $b_{0,1},\;d_{0,1},\;\phi_{0,1}$
satisfying the conditions
\bea 2b_0+2pd_0+af_0=0,\qquad 2b_1+2pd_1+af_1=\alpha \eea
and the constraint
\bea\label{ceDW0} 2 p\alpha d_1 =\left(ap d_1+ f_1\right)^2. \eea

\subsection{Standard solution}

Consider the case $\Delta \neq 0$. Without loss of generality we can set
\bea d_0=f_0=0. \eea Assuming now \bea \alpha=d_1=f_1=0
\eea we find that Eq. (\ref{Gwall}) reduces to \bea G=-2\ln|q\tau|, \eea
with \bea q^2 = \Delta b^2/4, \eea  and the solution reads
\bea\label{solDW0}
e^{2A}&=&\cF^{-2}|q\tau|^{\frac{4(p+1)}{\Delta p}},\nn\\
e^{2B}&=&e^{2D}=|q\tau|^{\frac4{\Delta p}},\nn\\
e^{a\phi}&=&|q\tau|^{-\frac{2a^2}{\Delta}}.\eea It has three
special points: $\tau=0,\;\pm\infty$. Consider the behavior of radial
null geodesics and the scalar curvature  in their vicinity. The
affine parameter in terms of $\tau$ will read  \bea
\lambda\sim|\tau|^{\frac{a^2+2/p}{\Delta}}, \eea while the curvature
scalar is \bea R\sim|\tau|^{-\frac{2a^2}{\Delta p}}. \eea

One can see that  for $\Delta>0$ the infinities $\tau=\pm\infty$
are at an infinite affine distance and the curvature scalar is
zero there. At $\tau=0$, which is at a finite affine distance, $R$
diverges (for $a\neq0$). To exclude the singularity we then have
to cut the solution at some finite $\tau$ introducing a material
brane.

For $\Delta<0$ the scalar curvature is zero at $\tau=0$  (if
$a\neq0$), and diverges at   $\tau=\pm\infty$. The surface $\tau=0$ is
at an infinite affine distance, and one can cut the solution to
avoid singularities at $\tau=\pm\infty$.

Now, if the singularities are at a finite affine distance,
consider the following coordinate transformation \bea\label{trDW}
-\tau=
q^{-1}H^{\epsilon},\qquad H=c+m|y|,\qquad m= q^{\epsilon}, \eea
where $\epsilon=\pm 1$. Then for $c>0$ the singularity will be cut
out. The free parameter $\epsilon$ can be used as follows:

a) If $\tau=\pm\infty$ is at infinite affine distance ($\Delta > 0$),
we choose
$\epsilon=+1$ so that the region $\tau\in (-\infty,-c/m]$ maps to
the semi-axes $y\neq0$. The domain wall locates at $\tau=-c/m$.

b) If $\tau=0$  is at infinity ($\Delta < 0$), we choose in (\ref{trDW})
$\epsilon=-1$, in which case the region $\tau\in[-m/c,0)$ maps to
the semi-axes $y \neq 0$, the domain wall will then be at $\tau=-m/c$.

In terms of these coordinates we arrive at the standard  domain wall
solution of Ref. \cite{Be99}: \bea \label{ssol}
ds^2&=&H^{\frac{4\epsilon}{\Delta p}}(-dt^2+dx_1^2+\ldots+dx_p^2)+
\frac{m^2}{q^2}H^{\frac{4(p+1)\epsilon}{\Delta
p}+2(\epsilon-1)}dy^2,\\
e^{a\phi}&=&H^{-\frac{2a^2\epsilon}{\Delta}},\label{ssolp}\\
R&=&\frac{b^2}2\left(a^2+\frac{p+2}{p}\right)
H^{-\frac{2a^2}{\Delta}\epsilon}.\eea

\subsection{Black solution}

Now let us seek for a more general solution without naked
singularities, but possibly endowed with a horizon. Repeating the
previous analysis,  we see that in the case $\Delta \neq 0$ the horizons
correspond to $\tau=\pm\infty$. Let $\tau=-\infty$ be the event
horizon, then from regularity we get the conditions:
\bea\label{horDW} d_1=-\frac{\alpha}2,\qquad f_1=-\frac{a\alpha}{\Delta},
\eea and the constraint (\ref{ceqDW}) holds automatically. Taking into
account (\ref{horDW}), setting  $d_0=0$ by a rescaling of $\{x^i\}$,
and trading the two integration constants $f_0$ and $\alpha$ for
\bea
y_0 &=& q^{-1}\e^{-a\phi_0/2} \qquad \left(\phi_0 = -\frac{\Delta p}{2(p+1)}\,f_0\right)\,,\\
\mu &=& \alpha y_0^2\,,
\eea
we arrive at the solution:
\bea
\e^{2B}&=&\bigg(4\frac{y_0^2}{\mu^2}e^{\alpha\tau}\sinh^2(\alpha\tau/2)\bigg)^{\frac2{\Delta
p}}e^{\alpha\tau},\\
\e^{2D}&=&\bigg(4\frac{y_0^2}{\mu^2}e^{\alpha\tau}\sinh^2(\alpha\tau/2)\bigg)^{\frac2{\Delta
p}},\\
\e^{2\A}&=&\bigg(4\frac{y_0^2}{\mu^2}e^{\alpha\tau}\sinh^2(\alpha\tau/2)\bigg)^{\frac{2(p+1)}
{\Delta p}}e^{\alpha\tau},\\
\e^{a\phi}&=&\bigg(4\frac{y_0^2}{\mu^2}e^{\alpha\tau}\sinh^2(\alpha\tau/2)\bigg)
^{-\frac{a^2}{\Delta}}e^{a\phi_0}.
\eea

Passing to $y$ via the map
\bea\label{tauDW} \tau=\frac1{\alpha}\ln \left(1-\frac{
\mu}{y}\right)  \eea we obtain
\bea &&ds^2=\bigg(\frac{y}{y_0}\bigg)^{-\frac4{\Delta
p}}\left[-\left(1-\frac{\mu}{y}\right)dt^2+d\vec{x}^2\right]
+\bigg(\frac{y}{y_0}\bigg)^{-2\left(1+\frac{a^2}{\Delta}\right)}
\left(1-\frac{\mu}{y}\right)^{-1}dy^2,\\
&&e^{a(\phi-\phi_0)}=\bigg(\frac{y}{y_0}\bigg)^{\frac{2a^2}{\Delta}}.
\label{bdw}\eea This solution can be interpreted as a black domain
wall. For $\mu=0$ it reduces to the standard solution
(\ref{ssol})-(\ref{ssolp}) with $c=0$, which is identical with the
linear dilaton background locally. So we can regard the solution
(\ref{bdw}) in the same way as our previous solutions: a black domain
wall on the linear dilaton background. For $\Delta < 0$ ($b^2 < 0$),
the linear dilaton asymptotic region $y \to \infty$ is at infinite
affine distance, with the other asymptotic region $y=0$ corresponding
to a singularity hidden behind the horizon. On the contrary, for
$\Delta > 0$ ($b^2 > 0$), the spacetime ends at the point singularity
$y \to +\infty$ (bag-of-gold black domain wall), with the other
asymptotic region $y=0$ at infinite affine distance.

In the case $\Delta=0$ one finds the possible horizons at
$\tau=\pm\infty$ as well. Choosing as the horizon
$\tau=-\infty$, we find the regularity conditions on the parameters:
\bea b_1=\frac{\alpha}2,\qquad d_1=f_1=0. \eea Choosing $\alpha = 1$, and
transforming the radial coordinate to $y = \e^{\tau}$, the
corresponding black domain wall solution is (up to rescalings)
\bea
&&ds^2 = \e^{-\frac{b^2}{p}y}\bigg(-ydt^2 + d\vec{x}^2\bigg) +
\e^{-\frac{b^2(p+1)}{p}y}\,\frac{dy^2}y\,, \\
&&\e^{a\phi} = \e^{\frac{b^2(p+1)}{p}y}.
\eea
In this case
the affine parameter for null radial geodesics reads, in the
asymptotic region $y \to +\infty$,  \bea
\lambda\propto
\e^{-\frac{b^2(p+2)}{2p}y},
\eea so that, again, for $b^2 < 0$ the manifold extends to infinity,
with a singularity at $y \to -\infty$, while for $b^2 > 0$ it
ends at the point singularity $y \to +\infty$, with $y \to -\infty$ at
infinite affine distance.

\section{Relation to $p$-brane solutions with extra
parameters}

The full $p$-brane solution was first derived in Ref.
\cite{ZhZh99}, using a rather intricate integration method and a
gauge different from the present one. The solution contained four
independent parameters, but it remained unclear whether it was
free of naked singularities. Later this solution  was reproduced
and analyzed in \cite{BrMaOz01} where it was suggested  that extra
parameters may occur, then  it was rederived with different
variations and analyzed in a number of papers (see for example
\cite{MiOh04,LuRo04}). It was stated that the full (asymptotically
flat) solution has four independent parameters. However, all the
previous solutions have defects, such as naked singularities. In
this section we will show that to avoid naked singularities, one
should fix additional parameters, and that the actual number of
parameters in good solutions (i.e. asymptotically flat, satisfying
cosmic censorship and having regular horizon) reduces to two.

Let us show this for the solution of Ref. \cite{ZhZh99} (all the
other solutions are related to this solution by simple change of
notations). To match our notations with those of Ref.
\cite{ZhZh99} one should set $E=0$, $d_0=0,\; q=k$ and set the
dilaton to zero at infinity, or equivalently
\bea \label{72} \phi_0=0,\qquad\frac{\alpha^2(k-1)^2}{\Delta b^2}=\sinh^2(\frac{\alpha\tau_0}{2}).
\eea
Making the following change of variables in (\ref{Sol1}-\ref{Sol6})
\bea
\tau=-\frac{1}{2r_0^{k-1}}\ln
\left(\frac{1-(\frac{r_0}{r})^{k-1}}{1+(\frac{r_0}{r})^{k-1}}\right),\quad
\mbox{and}\quad
\cF=-r^k\left(1-\frac{r_0^{2(k-1)}}{r^{2(k-1)}}\right). \eea and
replacing our parameters by those of Ref. \cite{ZhZh99}
\bea
&&\beta=4r_0^{k-1},\qquad\coth{\frac{\alpha\tau_0}{2}}=c_3,\qquad d_1=-c_2r_0^{k-1},\nn\\
&&\phi_1=2c_1r_0^{k-1}-\frac{a(d-3)}{(k-1)}c_2r_0^{k-1},\qquad \frac{\alpha}{\beta}=\tilde{k},
\eea
we obtain the general asymptotically flat solution in the form
of Ref. \cite{ZhZh99}
\bea
\label{ZhZh}
ds^2=-f(r)\e^{2A(r)}dt^2+\e^{2A(r)}(dx_1^2+...+dx_p^2)+\e^{2B(r)}(dr^2+r^2d\Sigma_k^2),
\eea
where
\bea
f(r)&=&\left[\frac{1-(\frac{r_0}{r})^{k-1}}
{1+(\frac{r_0}{r})^{k-1}}\right]^{-c_2},\\
A(r)&=&\frac{(k-1)(ac_1+(1+\frac{a^2}{2(k-1)})c_2)}{\Delta (d-2)}h(r)-\nn\\
&&\qquad-\frac{2(k-1)}{\Delta (d-2)}\ln[\cosh(\tilde{k}h(r))+c_3\sinh(\tilde{k}h(r))],\\
B(r)&=&\frac{1}{k-1}\ln\left[1-\left(\frac{r_0}{r}\right)^{2(k-1)}\right]-\frac{2(a(k-1)c_1-ac_2)}{2\Delta (d-2)}h(r)+\nn\\
&&\qquad +\frac{2(p+1)}{\Delta (d-2)}\ln[\cosh(\tilde{k}h(r))+c_3\sinh(\tilde{k}h(r))],\\
\phi(r)&=&-\frac{(k-1)(2(p+1)c_1-ac_2)}{\Delta (d-2)}h(r)-\frac{2a}{\Delta}\ln[\cosh(\tilde{k}h(r))+c_3\sinh(\tilde{k}h(r))],
\eea
and
\bea
h(r)=\ln\left(\frac{1-(\frac{r_0}{r})^{k-1}}{1+(\frac{r_0}{r})^{k-1}}\right).
\eea

The constraint (\ref{constraint}) transforms into the condition
\bea
\label{const2}
\frac{4}{\Delta}\tilde{k}^2+c_1^2-\frac{1}{\Delta}
\left(ac_1+\frac{(k-1)c_2}{d-2}\right)^2-\frac{2k}{k-1}+\frac{(d-3)}{2(d-2)}c_2^2=0.
\eea
So the complete asymptotically flat solution has four independent parameters
$r_\pm,\,c_2$ and $\tilde k$, as was stated in \cite{ZhZh99}.

To see the full structure of the solution (\ref{ZhZh}) let us now
make the transformation from the isotropic coordinate $r$ to a
Schwarzschild-type coordinate $\tilde r$
\bea
\lb{ZZ} r=\tilde r\left(\frac{\sqrt{f_+}+\sqrt{f_-}}{2}\right)^{\frac{2}{k-1}},
\eea
where
\bea
f_\pm=1-\left(\frac{r_\pm}{\tilde r}\right)^{k-1},\qquad r_0^{k-1}=\frac14(r_+^{k-1}-r_-^{k-1}),
\qquad c_3=-\frac{r_+^{k-1}+r_-^{k-1}}{r_+^{k-1}-r_-^{k-1}}.
\eea
The curvature has the following form in terms of the coordinate $\tilde
r$
\bea
R&\sim&\frac{(k-1)^2}{\tilde{r}^{2k}}(f_+f_-)^{-\frac{k}{k-1}-\frac{\tilde{k}(p+1)}{\Delta(d-2)}}
\left(\frac{f_+}{f_-}\right)^{\frac{a(p+1)}{\Delta(d-2)}c_1-\frac{a^2}{2\Delta(d-2)}c_2}
[r_+^{k-1}f_+^{-\tilde k}-r_-^{k-1}f_-^{-\tilde k}]^{-\frac{2(p+1)}{\Delta(d-2)}-1}\times\nn\\
&&\times\left[\mbox{const}_1f_+^{-\tilde k}+\mbox{const}_2f_-^{-\tilde k}+
\frac{b^2(d-2k)}{(k-1)^2(d-2)}\right]. \eea
From this expression we
see that the solution has an initial singularity at the point $\tilde
r=0$ and, for generic values of parameters, may be singular at
$\tilde r=r_-$ and $\tilde r=r_+$. Therefore to get a physically
interesting solution (i.e. satisfying the principle of cosmic
censorship) we have to demand the point $r_+$ to be a regular
horizon. The condition of its non-degeneracy ($n=1$ in terms of
formula (\ref{hor2})) gives us the following constraint on parameters
\bea
2a(d-2)c_1-a^2(d-2)c_2-2p(k-1)c_2+4(d-2)\tilde k-\frac{2\Delta k
(d-2)}{k-1}=0,
\eea
using which we can rewrite (\ref{const2}) as
\bea
\frac{pk}{2(d-2)}(c_2+2)^2+\frac1{a^2(d-2)^2}(2k(p+1)+p(k-1)c_2-2(d-2)\tilde
k)^2=0.
\eea
This condition fixes two of the four parameters
\bea c_2=-2,\qquad \tilde k=1, \eea
and two independent parameters
remain, in accordance with the result in section 6.

Let us make some remarks on the solution obtained in Ref.
\cite{LuRo04}. It can be derived from our generic solution
(\ref{solution}-\ref{Sol6}) using the same gauge function as in
(\ref{gauge}) $(q=k)$
\bea
\tau=\frac2{\beta}\ln\left(\frac{1-(\frac{w}{r})^{q-1}}
{1+(\frac{w}{r})^{q-1}}\right)\equiv \frac2{\beta}\ln\frac{\tilde H_1}{H_1},\quad  \mbox{and}\quad
\cF=r^q\left(1-\frac{w^{2(k-1)}}{r^{2(k-1)}}\right)\equiv
r^qH_1\tilde H_1. \eea where \bea \beta=4w^{q-1}.
\eea
 Fixing the following set of parameters as
\bea
d_{0,1}=e_{0,1}=0,\qquad g_1=a\phi_1,\qquad  \frac{\alpha^2(q-1)^2}{\Delta b^2}=\sinh^2(\alpha\tau_0/2),
\eea
identifying the remaining parameters with those of Ref. \cite{LuRo04}
\bea
\chi=\frac{\Delta(d-2)}{q-1},\qquad\cosh^2\theta=\frac1{1-\e^{\alpha\tau_0}},\qquad\delta=-\frac{2\phi_1}{\beta}
\eea
and replacing the
combinations $-(\alpha+a\phi_1)/\beta$ and $-(\alpha-a\phi_1)/\beta$ by $\alpha$ and $\beta$ respectively we
obtain the solution in the form of \cite{LuRo04}
\bea
&&ds^2=(H_1\tilde H_1)^{\frac2{k-1}}F_1^{\frac{4(p+1)}{\chi(q-1)}}
(dr^2+r^2d\Sigma_q^2)+F_1^{-\frac4{\chi}}(-dt^2+dx_1^2+\ldots+dx_p^2),\\
&&\e^{a\phi}=\left(\frac{H_1}{\tilde H_1}\right)^{a\delta}
F_1^{\frac{-2a^2(d-2)}{\chi(q-1)}},
\eea
where
\bea
F_1=\cosh^2\theta\left(\frac{H_1}{\tilde H_1}\right)^\alpha-\sinh^2\theta \left(\frac{\tilde H_1}{H_1}\right)^\beta.
\eea
This solution has three independent
parameters: $\delta,\,w$ and $\theta$. However, as one can easily
check (by the transformation to the Schwarzschild-type coordinate
(\ref{ZZ})), this solution has a horizon at $r_+$, which
generically is not regular. Moreover, the parameters of the
solution \cite{LuRo04} cannot be adjusted to give a regular horizon,
because, as was pointed out in \cite{LuRo04}, this solution
corresponds to $c_2=0$ of Ref.\cite{ZhZh99}, but we have found
earlier in this section that the regularity of the horizon of
Ref.\cite{ZhZh99} implies
\bea c_2=-2. \eea Indeed,  one can check that, after transforming
to Schwarzschild type coordinates and imposing the regularity
conditions (\ref{parhor}) one obtains a complex value for the
parameter $\delta$, which is not allowed by the construction of
\cite{LuRo04}. Therefore within the parameter space of the
solution of Ref. \cite{LuRo04} we do not find solutions satisfying
the conditions(\ref{parhor}) necessary for regularity. The above
reasoning applies to the first branch of the general solution of
Ref. \cite{LuRo04} (the one with the real functions $H_1$ and
$\tilde{H_1}$). We suppose that their second branch with imaginary
$H_1$ and $\tilde{H_1}$ corresponds to imaginary values of our
parameters $\alpha$ and/or $\beta$. In this case we did not find
solutions with a regular horizon either.

\section{Conclusions}

The purpose of this paper was to clarify the meaning of special
points in the general supergravity $p$-brane solution with a single
 charge and to find all possibilities for  the values of
parameters when the solution is free from naked singularities,
though not necessarily asymptotically flat. We have found that
there are three options depending on the value of the parameter
$\tau_0$ marking the position of the (generic) singularity. For
$\tau_0>0$, the singularity is hidden inside the event horizon,
and the asymptotically flat solution coincides with the standard
black $p$-brane of \cite{DuLuPo96}. For $\tau_0<0$, the solution
has a regular horizon, but is singular at spatial infinity. For
$\tau_0 = 0$, the regular black solution is not asymptotically
flat, and constitutes a delocalized generalization of the black
$p$-brane on the linear dilaton background discussed in
\cite{ClGa04}. We also presented an explicit identification of the
solution by Zhou and Zhu in terms of our solution, thus clarifying
the nature of ``the extra parameters''. In addition we compared
our solution with the recent results by Lu and Roy. Our general
conclusion is that no $p$-brane solutions possessing a regular
event horizon and not plagued with naked singularities exist,
other than the standard asymptotically flat black branes and the
black branes on the linear dilaton background.

\vspace{0.2in}

\acknowledgments{This work was supported in part by RFBR grants
02-04-16949 and 04-02-16538.}

\clearpage

\footnotesize \psfrag{0}{\rm$\tau=0\;(r=\infty)$}
\psfrag{t0}{\rm$\tau=\tau_0\;(r=r_0)$}
\psfrag{tpi}{\rm$\tau=+\infty\;(r=r_-)$}
\psfrag{tmi}{\rm$\tau=-\infty\;(r=r_+)$}
\psfrag{itpi}{\rm$\tau=+\infty+\frac{i\pi}{\alpha}\;(r=r_-)$}
\psfrag{itmi}{\rm$\tau=-\infty+\frac{i\pi}{\alpha}\;(r=r_+)$}
\smallskip
\FIGURE{\hspace{-1cm}\epsfig{file=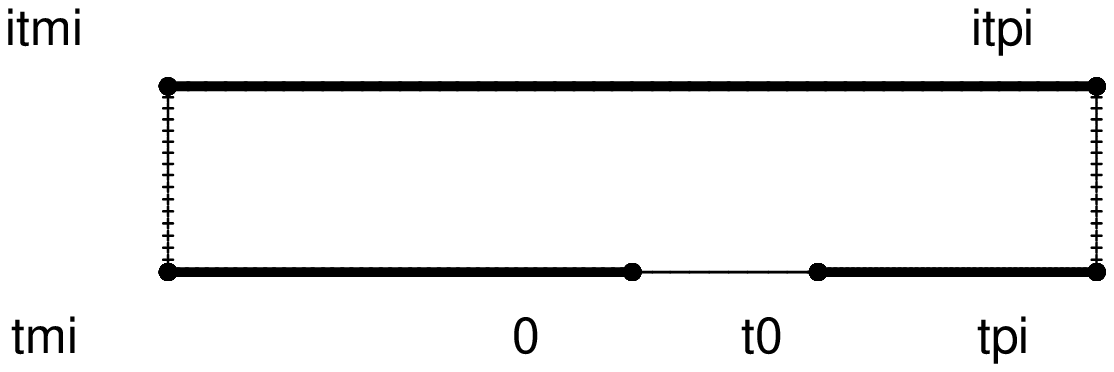,width=14cm}
        \caption{Complex plane of $\tau$ for an asymptotically flat
solution. In this case $\tau_0$ is strictly positive and
corresponds to the central singularity. The real axis corresponds
to regions outside the event horizon and inside the inner horizon.
Between the horizons $\tau$ has an imaginary part (upper line).
The dependence $\tau(r)$ corresponds to the map
(\ref{tauxi},\ref{rxi}).}}

\psfrag{0}{\rm$\tau=\tau_0=0\;(r=\infty)$}
\psfrag{t0}{\rm$\tau=\tau_0\;(r=\infty)$}
\FIGURE{\hspace{-1cm}\epsfig{file=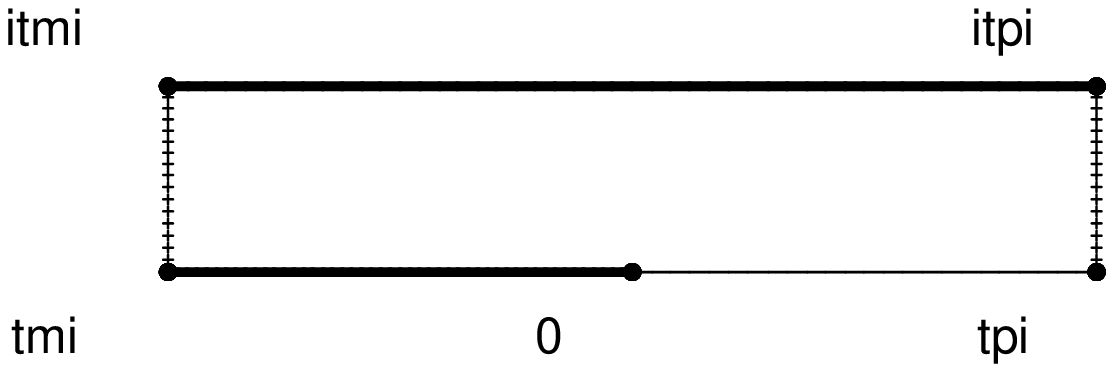,width=14cm}
        \caption{Complex plane of $\tau$ for the solution with the linear
dilaton asymptotic. In this case $\tau_0$ is non-positive and to
get a solution without naked singularities one has to take the
limit $\tau_0\to -0$.}}


\begin{thebibliography}{99}

\bibitem{HoSt91} G.~T.~Horowitz and A.~Strominger,
{\it Black strings and p-branes}, \npb{360}{1991}{197-209}.

\bibitem{Gu92} R.~G\"uven, {\it Black $p$-brane solutions of $D=11$
supergravity theory}, \plb{276}{1992}{49}.

\bibitem{DuLu94} M.~J.~Duff and J.~X.~Lu, {\it Black and Super $p$-Branes in
Diverse Dimensions}, \npb{416}{1994}{301} \hepth{9306052}.

\bibitem{LuPoSeSt95}  H.~L\"u, C.~N.~Pope, E.~Sezgin, K.~S.~Stelle
{\it Stainless super p-brane}, \npb{456}{1995}{669-698}
\hepth{9508042}.

\bibitem{LuPo95}  H.~L\"u and C.~N.~Pope,
{\it p-brane solitons in maximal supergravities},
\npb{465}{1996}{127-156} \hepth{9512012}.

\bibitem{LuPoXu95}  H.~L\"u,  C.~N.~Pope, and K.~W.~Xu,
{\it Liouville and Toda Solitons in M-theory},
\mpla{11}{1996}{1785-1796} \hepth{9604058}.

\bibitem{DuKhLu95} M.~J.~Duff, R.~R.~Khuri and J.~X.~Lu,
{\it String Solitons}, \pr{259}{1995}{213} \hepth{9412184}.

\bibitem{DuLuPo96}  M.~J.~Duff, H.~L\"u and C.~N.~Pope,
{\it The Black Branes of M-theory}, \plb{382}{1996}{73}
\hepth{9604052}.

\bibitem{LuPoSeSt96}
H.~L\"u, C.~N.~Pope, E.~Sezgin, K.~S.~Stelle,  {\it Dilatonic
p-brane solitons}, \plb{371}{1996}{46-50} \hepth{9511203}.

\bibitem{St98} K.~S.~Stelle, {\it BPS Branes in Supergravity}, \hepth{9803116}.

\bibitem{GaRy98}  D.~V.~Gal'tsov and O.~A.~Rytchkov, {\it Generating Branes
via Sigma-Models},  \prd{58}{1998}{122001} \hepth{9801160}.

\bibitem{IvMe01} V.~D.~Ivashchuk and V.~N.~Melnikov,
{\it Exact solutions in multidimensional gravity with
antisymmetric forms, topical review}, \cqg{18}{2001}{R87-R152}
\hepth{0110274}.

\bibitem{Du99} M.~J.~Duff, {\it Tasi lectures on branes,
black holes and anti-de Sitter space}, \hepth{9912164}.

\bibitem{ClGa04}
G.~Cl\'ement, D.~Gal'tsov and C.~Leygnac, {\it Black branes on the
linear dilaton background}, \prd{71}{2005}{084014} \hepth{0412321}.

\bibitem{GaLeCl04}  D. Gal'tsov, J. Lemos and G. Cl\'ement, {\it
Supergravity $p$-brane reexamined: extra parameters, uniqueness
and topological censorship}, \hepth{0403112}.

\bibitem{ClGaLe02} G. Cl\'ement, D. Gal'tsov and C. Leygnac,
{\it Linear dilaton black holes}, \prd{67}{2003}{024012}.

\bibitem{ClLe04}
G. Cl\'ement, C. Leygnac,  {\it Non-asymptotically flat, non-AdS
dilaton black holes}, \prd{70}{2004}{084018}.

\bibitem{Le04} C. Leygnac, {\it Non-asymptotically flat black
holes/branes}, \grqc{0409040}.

\bibitem{ItMaSoYa98} N. Itzhaki, J.M. Maldacena, J. Sonnenschein,
and S. Yankielowicz, {\it Supergravity and The Large N Limit of
Theories With Sixteen Supercharges}, \prd{58}{1998}{046004}.

\bibitem{Be99}
K. Behrndt, E.  Bergshoeff, R. Halbersma, J.  P. van der Schaar,
{\it On Domain-Wall/QFT Dualities in Various Dimensions}, \cqg{16}{1999}{3517-3552}\\
E. Bergshoeff, R. Halbersma {\it On Domain-Wall/QFT Dualities in
Various Dimensions}, \hepth{0001065}. Talk given at International
Workshop on Physical Variables in Gauge Theories, Dubna, Russia,
21-25 Sep 1999.

\bibitem{BoSkTo99} H.J. Boonstra, K. Skenderis, P.K. Townsend
{\it The domain-wall/QFT correspondence}, \jhep{01}{1999}{003}.

\bibitem{MaSt97} J. Maldacena and A. Strominger,
{\it Semiclassical decay of near extremal fivebranes},
\jhep{12}{1997}{008}.

\bibitem{ZhZh99}
B.~Zhou and C.-J.~Zhu, {\it The complete black brane solutions in
d-dimensional coupled gravity system}, \hepth{9905146}.

\bibitem{BrMaOz01}
P.~Brax, G.~Mandal and Y.~Oz, {\it  Supergravity description of
Non-BPS branes}, \prd{63}{2001}{064008} \hepth{0005242}.

\bibitem{OhYok02}
Ohta and Y. Yokono, {\it Gravitational approach to tachyon
matter}, \prd{66}{2002}{125009}.

\bibitem{LuRoyDel04}
J. X. Lu and S. Roy, {\it Delocalized, non-SUSY $p$-branes,
tachyon condensation and tachyon matter},  \jhep{11}{2004}{008}.

\bibitem{KoAsMa04}
S. Kobayashi, T. Asakawa, and S. Matsuura, {\it Open string
tachyon in supergravity solution}, \hepth{0409044}.

\bibitem{Se99}
A.~Sen, {\it Non-BPS states and branes in string theory},
\hepth{9904207}.

\bibitem{BerVecFra00}
Bertolini, P. Di Vecchia, M. Frau, A. Lerda, R. Marotta and R.
Russo, {\it Is a classical description of stable non-BPS
D$p$-branes possible?}, \npb{590}{2000}{471}.

\bibitem{Frau:2000gk}
M.~Frau, A.~Liccardo and R.~Musto, {\it The geometry of fractional
branes}, \npb{602}{2001}{39} \hepth{0012035}.

\bibitem{Alberghi:2001fy}
G.~L.~Alberghi, E.~Caceres, K.~Goldstein and D.~A.~Lowe, {\it
Stacking non-BPS D-branes}, \plb{520}{2001}{361} \hepth{0105205}.

\bibitem{Bertolini:2002de}
M.~Bertolini, T.~Harmark, N.~A.~Obers and A.~Westerberg, {\it
Non-extremal fractional branes},\npb{632}{2002}{257}
\hepth{0203064}.

\bibitem{ChGaGu02}
C.-M.~Chen, D.~V.~Gal'tsov and M.~Gutperle {\it S-brane solution
in supergravity theories}, \prd{66}{2002}{024043} \hepth{0204071}.

\bibitem{Lu}
J.~X.~Lu, {\it ADM Masses for Black Strings and p-Branes},
\plb{313}{1993}{29} \hepth{9304159}.

\bibitem{Argurio}
R.~Argurio, {\it Brane Physics in M-theory}, \hepth{9807171}.

\bibitem{BY} J.D. Brown and J.W. York, {\it Quasilocal Energy and
Conserved Charges Derived from the Gravitational Action},
\prd{47}{1993}{1407}.

\bibitem{HaHo95} S.W. Hawking and G.T. Horowitz,
{\it The Gravitational Hamiltonian, Action, Entropy, and Surface
Terms}, \cqg{13}{1996}{1487}.

\bibitem{ChNe98}
C.M. Chen, J.M. Nester, {\it Quasilocal quantities for GR and
other gravity theories}, \cqg{16}{1999}{1279-1304}.

\bibitem{Bo00} I.S.N. Booth, {\it A Quasilocal Hamiltonian for
Gravity with Classical and Quantum Applications}, Ph.D Thesis
\grqc{0008030}.

\bibitem{MiOh04}
Y.-G.~Miao and N.~Ohta, {\it Complete Intersecting Non-Extreme
p-Branes}, \plb{594}{2004}{218} \hepth{0404082}.

\bibitem{LuRo04}
J.~X.~Lu and S.~Roy, {\it Static, non-SUSY $p$-branes in diverse
dimensions}, \hepth{0408242}.



\end{thebibliography}
\end{document}